\documentclass[conference]{IEEEtran}

\usepackage{graphicx}
\usepackage{amsmath}
\usepackage[T1]{fontenc}
\usepackage{times}          
\usepackage{url}
\usepackage[hidelinks]{hyperref}

\begin{document}

\title{Insights on Student Learning from Live Classroom Polls: More Than Right
or Wrong}

\author{%
\IEEEauthorblockN{Rohit Sharma\IEEEauthorrefmark{1}, Pavani Ayinampudi\IEEEauthorrefmark{2}, Aditya B.M.V.\IEEEauthorrefmark{2}, Jinal Gupta\IEEEauthorrefmark{2},\\
Prakash Hegade\IEEEauthorrefmark{2}, Sakshi Sharma\IEEEauthorrefmark{1}, Meenakshi V\IEEEauthorrefmark{1}, SRS Iyengar\IEEEauthorrefmark{1}}
\IEEEauthorblockA{\IEEEauthorrefmark{1}Indian Institute of Technology Ropar, Rupnagar, Punjab 140001, India}
\IEEEauthorblockA{\IEEEauthorrefmark{2}ANNAM.AI, Rupnagar, Punjab, India}%
}

\maketitle

\begin{abstract}
A live classroom poll is usually read through its answer key, as a count of participation and correctness. Yet the same record holds more before any key is fixed, namely who answers early and how the class divides across the options. Whether these key-free signals are stable, and what they reveal, has not been examined at the scale of a real course. We study them in a live deployment of 340,668 submissions across 603 polls in 47 sessions from roughly 2,400 learners, reading each poll for a learner's response order and for how much the class divides. A single further step brings in the author-designated key, on items with a settled answer, to interpret the divided ones. A learner's response order is a stable individual signature. It reproduces at a corrected split-half reliability of 0.92, holds at 0.87 within single sessions, and is unrelated to whether the learner is correct. Read without a key, about one poll in five does not reach a clear majority, and division marks the questions that split the class. On the settled-answer items the majority is itself wrong on 49 of 358. The key-free measure flags most of these, though division alone does not separate a collectively wrong answer from a merely hard one. The two readings are largely independent, meeting at one point where early responders anticipate the eventual majority only on questions the class has nearly settled. Together they map the learners and questions of a session without a key.
\end{abstract}

\vspace{2mm}

\section{Introduction}
A teacher wants to know whether a class is following along, and in a large class, that is hard to judge. Real-time polling grew out of this need and is extensively used through clickers, web-based response tools, and video-conferencing polling software, with meta-analytic evidence pointing to genuine but contingent benefits (Caldwell, 2007; Kay \& LeSage, 2009; Hunsu et al., 2016). We use these polls to infer the level of understanding of a class. When an instructor pauses to ask a question, the responses arrive within seconds, and the inferences about the students in the room are made based on two numbers: how many students answered and when an answer key exists, how many were right. It is well established that retrieving an answer, even under low stakes, consolidates memory (Roediger \& Karpicke, 2006), and a poll returns feedback that can redirect teaching while the lesson is still underway (Black \& Wiliam, 1998). Both of the usual readings, the participation count and the correctness rate, treat the poll as a tally.

That tally leaves out much of what a poll response shows. Before any answer key is applied, the same record already shows who answers early and who answers late, and where a class agrees and where it splits. These are ordinary properties of the responses, visible the moment they come and needing no correct answer. Yet they disappear once a poll is read only as how many took part and how many were right. A class that agrees strongly and one that divides evenly can return the same participation count, but a tally hides how the class behaved and what kind of question was asked.

Whether that discarded description can be trusted is not obvious. A pattern visible in the data need not be stable: a learner's apparent tendency to answer early might reflect a passing mood rather than a durable trait, and a question's apparent division might be an artefact of a single session rather than a property of the question. If such signals are unstable, reading them would mislead more than it can inform. If they are stable, they would describe a live class in ways a correctness score cannot, and doing so while the session is still under way, when the description can still shape what happens next. Which of these holds is an empirical matter, and it is not settled.

Research on classroom response systems has treated the poll as an assessment moment, asking what a correct answer reveals rather than what the response record reveals before an answer key is applied. And the datasets that might have settled the matter have been very small, a handful of sessions or a single class too limited to separate a stable signal from noise. What a live poll reveals about a class before any key is applied, whether those signals are stable enough to be read as descriptions, and whether a reading of the learners and a reading of the questions carry the same information, remain open at the scale of a real course. These are the questions this paper sets out to answer.

The rest of the paper is organised as follows. Section II presents the background study, Section III describes the methodology and methods, Section IV presents the data collection and analysis, Section V presents the results and discussion and Section VI concludes.

\section{Background Study}
Research on classroom response systems has been done over the last two decades, with generally favourable findings. There is a gain in participation, attention, and timely feedback as per the reviews of clickers (Caldwell, 2007). Broader reviews of audience-response systems report the same (Kay \& LeSage, 2009), while a meta-analysis of more than fifty studies calls these gains genuine but modest and contingent on use (Hunsu et al., 2016). More recent web and mobile platforms carry the tradition onto learners' own devices, with comparable engagement gains (Wang \& Tahir, 2020). In the peer-instruction tradition the most useful items are those that split the room, because a split reveals a live disagreement worth surfacing (Mazur, 1997). The sharpest divisions yield the largest gains from peer discussion (Crouch \& Mazur, 2001), and the discussion itself drives the improvement (Smith et al., 2009). This tradition treats the poll as an assessment moment, reading participation and correctness and using a split as a cue for the instructor rather than a quantity to be measured.

A separate group of people studied response latency in computer-based testing, where the interval between item presentation and answer is a well-developed signal. Hierarchical frameworks jointly model speed and accuracy (van der Linden, 2007), and reviews survey how response times inform difficulty, engagement, and rapid guessing (De Boeck \& Jeon, 2019). Response-time-effort measures flag disengaged rapid responding in low-stakes settings (Wise \& Kong, 2005), and classic accounts link the time taken to the process of arriving at an answer (Wickelgren, 1977). This work measures latency from a recorded launch instant. It is long established that mental speed is a reliable individual difference (Jensen, 2006). Learning analytics likewise turns fine-grained traces into learner-level indicators (Siemens \& Baker, 2012), whose reliability is assessed through split-half and intraclass correlation (Shrout \& Fleiss, 1979).

The item side uses item analysis in educational measurement, traditionally centred on difficulty and discrimination, both defined against a scored key (Crocker \& Algina, 1986). These classical indices (Ebel \& Frisbie, 1991) are powerful inside a well-keyed test but presuppose a settled key. The closest prior work to a key-free reading is distractor analysis, where functioning distractors reveal what test-takers believe (Haladyna \& Downing, 1989), reaffirmed in later item-writing syntheses (Haladyna et al., 2002). The pattern of chosen distractors can point to specific misconceptions (Gierl et al., 2017), which links to instruments whose wrong options match identifiable alternative conceptions (Hestenes et al., 1992). Standard summaries of categorical spread, including normalised Shannon entropy and agreement measures for ordered categories (van der Eijk, 2001), offer complementary views of a distribution. Both sides sit inside the formative-assessment tradition, where an in-class question generates evidence that shapes the next move (Black \& Wiliam, 1998; Wiliam, 2011), and where separating genuine difficulty from item ambiguity is a long-standing concern handled through expert review and cognitive interviews (Willis, 2005).

The collective answer a poll produces has its own literature on social influence and the accuracy of aggregated crowd judgments, which bears on how early responders anticipate a consensus. The conditions for accurate aggregation of independent judgments are well known (Surowiecki, 2004). So is how exposure to others weakens the independence that makes crowds wise (Lorenz et al., 2011), and how public response order shapes individual answers (Asch, 1956).

Across these traditions the poll is read either through a settled key in item analysis or as a cue for discussion in peer instruction. The response record before any key is fixed is left largely unexamined, and three gaps follow. Whether a learner's response order is stable has not been established for live polling. What a poll's answer distribution reveals without a key has not been characterised at scale. Whether the two readings are related remains unknown.

\section{Methodology and Methods}

\subsection{Research Questions}
This study is organised around one main research question: what can a live classroom
poll reveal about the learners and the questions before its answer key is applied? We
decompose it into three sub-questions, two for the separate readings and one for their
relationship:
\begin{itemize}
\item \textit{RQ1.} What is the implication of ranking learners by how early they respond to a poll?
\item \textit{RQ2.} What does the spread of answers across the options reveal about a poll, before its key is applied?
\item \textit{RQ3.} What is the relationship between the reading of the learners and the reading of the questions?
\end{itemize}
RQ1 reads the learners and RQ2 the questions, each without a key. RQ3 relates the two, where the paper's central claim lies.
All constructs and every association we report are descriptive, drawn from a
single observational deployment, and we make no causal claim.

\subsection{The Response-Order Construct}
For a given poll, the respondents are the learners who submitted an answer. We sort
them by submission time and rank each as a percentile between zero and one, so the
earliest respondent falls near zero and the latest near one. Ties use the
average-rank method, so simultaneous submissions share a value. A learner's
response-order score is the average of these within-poll percentiles across all
polls they answered. A score near zero describes a learner who tends to answer
first, and one near the top a learner who tends to answer last.

The essential property is that the percentile is computed inside each poll. Whatever
instant a poll was launched is the same for everyone answering it, so it adds an
identical offset to every submission time and cancels when respondents are ranked
against one another. The measure is therefore a relative order, invariant to the
unknown launch instant, not a reaction time. For description only we retain the
seconds each response arrived after the first on its poll, used to characterise how
tightly polls fill and, in the crossing, how quickly each item resolves. The
reliability analyses restrict to learners with at least ten answered polls. The
median eligible learner answered 87. Comparing response order across learners
assumes a global timer, which the within-session check below probes.

\subsection{Key-Free Disagreement Measures}
For each cognitive fixed-choice poll we compute four quantities from its response
distribution, none of which uses a correct answer. The majority-option share is
the fraction of responses going to the single most chosen option, high when the
class agrees and low when it divides. The second-option share is the fraction
going to the second most chosen option. The normalised Shannon entropy captures
overall spread, scaled so that a perfectly even split is one and a unanimous
response is zero. The number of options records how many distinct answers
appeared. We code item type from the option structure, with exactly true and false as true or
false, three to eight fixed options as multiple-choice, and a very large number of
distinct answers as open response, which is excluded. We flag an item as contested
when the second-option share is at least 30 percent, and more strictly at least 40
percent. Read without a key, the flag marks only that a large block settled on one
competing option, not whether that block is right or wrong, so we treat it purely as a measure of division and make no claim about shared ideas. To
turn these continuous measures into interpretable groups, we cluster the items with
$k$-means, a standard clustering method, on three standardised features, the majority-option share, the
second-option share, and the normalised entropy. We fix the number of groups at three, chosen for interpretation rather than by a fit criterion, to separate
settled, moderately divided, and strongly divided items. For the two-option items
that make up most of the corpus these three features are all functions of one
quantity, the size of the split, so for those items the clustering orders items by
how divided they are rather than uncovering independent structure.
Because disagreement is not difficulty, we also reviewed the twenty most divided
items by text, coding each as genuine conceptual difficulty, ambiguous wording, or a
mislabelled opinion prompt, with a one-line rationale. This is a transparent author
judgment on a sample, not a validated instrument.

\subsection{Crossing the Two Axes}
The crossing joins the per-submission timing of the person axis to the per-poll
disagreement of the item axis on the cognitive polls, matching each poll across the
two analyses by session and order within the session. We ask four questions. Whether
items fill more slowly when more divided: for each poll we relate the time to ninety
percent of responses, the first-to-last span, and the median seconds after the first
to its disagreement measures and cluster, controlling for respondents. Whether early
responders match the eventual majority more on some items: we take the
majority-alignment gap between the earliest and latest thirds and compare it across
clusters. Whether the competing block on contested items fills late: on flagged
items we compare the mean response-order percentile of majority choosers against
strongest-competitor choosers. And whether response order relates to majority
selection differently on divided than settled items, at the learner level. All
associations are reported with significance and read descriptively.

\section{Data Collection and Analysis}

\subsection{Study Context and the Poll System}
The data derive from a live polling deployment during an online summer internship
programme delivered over several weeks across 47 recorded sessions. The programme
is oriented towards acclimatising and motivating incoming interns rather than
assessing mastery, and during each session the instructors ran frequent polls
covering mood checks, logistics, icebreakers, and content questions, which students
answered live on their own devices. This orientation matters for
interpretation. It explains why most content items are short two-option questions,
and throughout we separate the findings likely to hold beyond a setting like this
from those that may be specific to it. The platform exported,
for each poll, the full list of responses with the chosen option text and a
submission timestamp at one-second resolution.

\subsection{Data Sources and the Analysis Corpora}
Across the 47 sessions the system launched 603 polls and captured 340,668 submissions
from 2,422 distinct learner identities, each keyed on the join email so a learner is
tracked across the programme. The two axes use overlapping but not identical slices,
for principled reasons. The person axis is defined for any poll, since response order
needs only more than one respondent, so it uses all polls. From reliability
estimation we exclude only three unsuitable parts, one special session outside the
schedule, one late session with two participants, and one early session with a single
poll, leaving 596 polls and 2,378 learners. The item axis concerns agreement about
content, meaningless for a mood or icebreaker poll, so it restricts to the cognitive
subset. After the same exclusions and setting aside two open-response items, 505
cognitive fixed-choice polls remain, a median of 608 responses each. Of these, 476
are true or false and only 29 multiple-choice, so we report the two types separately
throughout. The crossing joins person timing to item disagreement on these 505 polls.
The two analyses' independent response counts agree exactly on 498 and to within two
on all but one.

One measurement fact shapes the person axis. The export records when each response was
submitted but not when each poll was launched, which rules out reaction time and
directs us to relative response order, defined from the respondents' submission
timestamps so that the unknown launch instant cancels.

\subsection{Descriptors and Data Protection}
Two descriptors accompany each cognitive item, used only as grouping variables. The
first is a cognitive level from Bloom's taxonomy, inferred from the surrounding
lecture, on which the corpus is dominated by lower-order items. The second, for the
key-referenced step, is the correct answer designated by each question's author at
creation, restricted to items that carry such a settled answer. A voluntary
end-of-programme self-report survey on a five-point scale is joined through the same
hashed identifier and supplies the few items used only to interpret the
response-order signature in Section~V. No raw identifier appears in any result.
Identity is hashed with a local salt before any computation beyond parsing, so every
reported quantity aggregates over anonymous keys and no individual can be
identified. Example items are paraphrased by topic. Informed consent for research
use was obtained from all participating students before data collection.

\subsection{Analysis}
The stability of the response-order score is quantified three ways so that no single
choice carries the trait claim. For split-half reliability we randomly partition each
learner's polls into two halves, correlate the half scores across learners, and apply
the Spearman-Brown correction, over 200 random splits, reporting a mean and interval
under both Spearman and Pearson. We also report the intraclass correlation, how much
of the score difference is between rather than within learners, in single-poll and
averaged forms. To rule out cross-day clock drift we repeat the split-half inside
single sessions, where the day and shared clock are fixed. On the item axis we
compute the disagreement measures and clusters above. As a single key-referenced
step, the one place a key enters, we use the correct answer that each question's
author designated at creation, stored with the item. We restrict it to the 358 items
carrying such an answer, excluding opinion prompts and items without a recorded key,
and record whether the majority option is correct. This step interprets the contested
items but does not define them, and the key plays no part in the disagreement
measures. The crossing uses rank correlations throughout, including one holding poll
size constant, a test of whether the three clusters differ in fill speed
(Kruskal-Wallis), and a per-poll test of the early-minus-late gap. An online
supplement gives four tables covering corpus composition, reliability by minimum-poll
threshold, the three cluster centroids, and the key-referenced check of collective
errors by disagreement band.\footnote{Online supplement:
\url{https://osf.io/e2t6q/?view_only=7226445db09f462a8ffbfa6f75eb8cab}}

\section{Results and Discussion}

\subsection{Results}
\textit{A stable response-order signature (RQ1).} Relative response order is highly reliable. Across the 1,785 learners with at least ten answered polls, over 200 random splits the split-half
correlation between a learner's two half scores is 0.86 before correction and 0.92
after the Spearman-Brown correction, with a 95 percent interval of 0.91 to 0.93 and
effectively the same value under Pearson. The intraclass correlation separates two
components: the single-measure value is 0.21, so any one poll is a noisy read on a
learner, while the average-measure value, over many polls, is 0.98. That gap is the
point of the construct: a single response order is noisy, but averaging over dozens of
polls yields a very stable index. The stability is not clock drift across days:
repeating the analysis inside single sessions, where the day and shared clock are
fixed, gives 0.87 over 15,874 learner-session pairs, barely moving the number, as
expected if the signature belongs to the learner rather than the recording system.

\begin{figure}[t]
\centering
\includegraphics[width=\columnwidth]{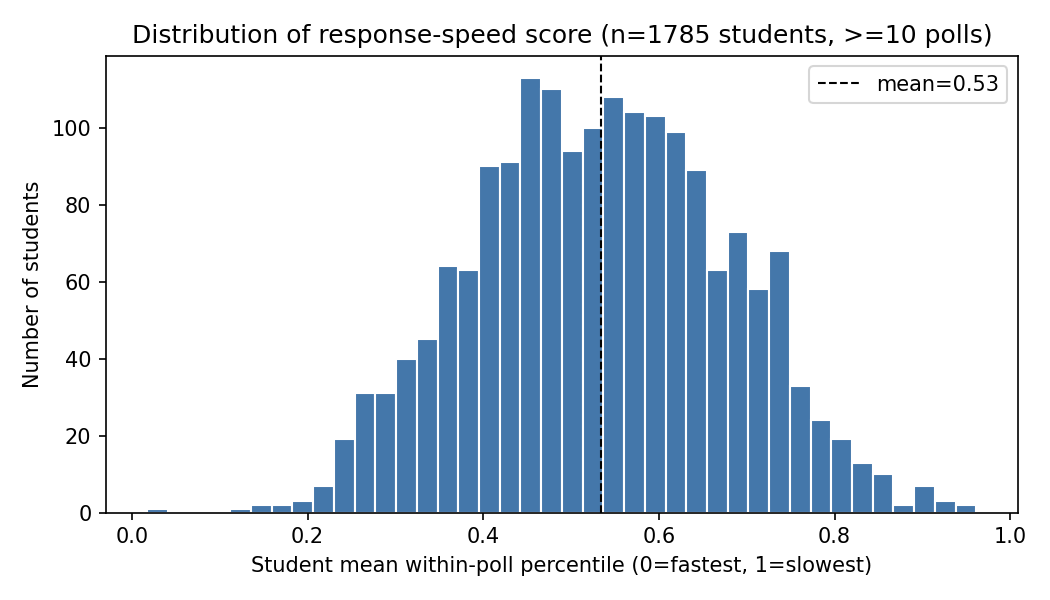}
\caption{Distribution of the response-order score across learners with at least ten
polls. The mass spreads from a fast decile near the front of the order to a slow
decile near the back.}
\label{fig:spread}
\end{figure}

\begin{figure}[t]
\centering
\includegraphics[width=\columnwidth]{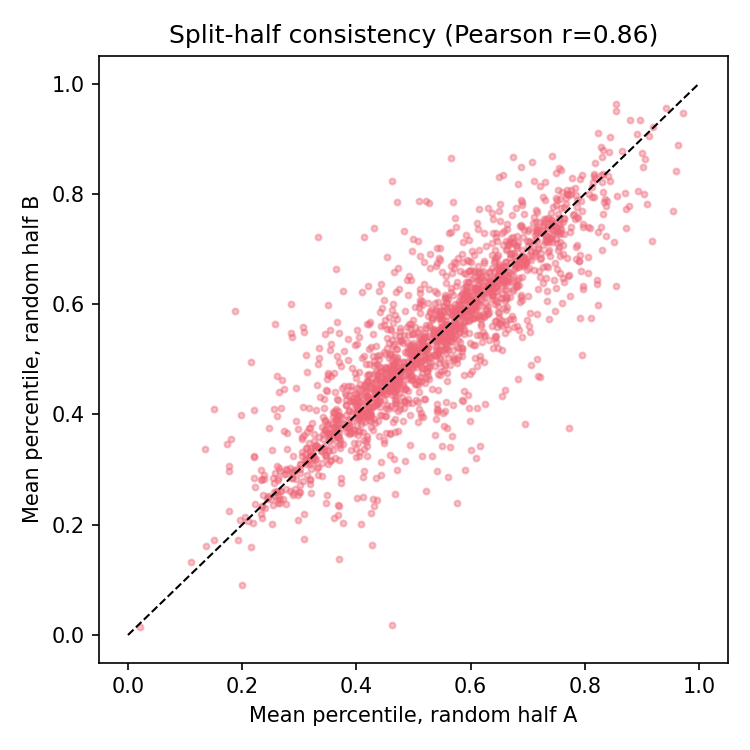}
\caption{Split-half consistency of the two half scores across learners. Each
learner's two half scores track the diagonal closely, with little off-diagonal
scatter.}
\label{fig:splithalf}
\end{figure}

The signature spreads learners widely (Fig.~\ref{fig:spread}). Among learners with at
least ten polls the mean score is 0.53 with a standard deviation of 0.14, and the
tenth and ninetieth percentiles sit at 0.35 and 0.73: the fastest tenth average a
within-poll percentile of 0.29, near the front, the slowest 0.78, well towards the
back. This ordering lives inside a near-instant window, since the median poll spans 13
seconds from first response to last and reaches ninety percent of its responses in a
median of 9 seconds. A difference of a few positions is thus a second or two of
wall-clock time, a fine-grained behavioural distinction rather than a coarse split
between attentive and absent learners. Because a response is a single option selection
under a fixed window, this ordering reflects when a learner chooses to commit, not how
fast they type or their connection replies, and a learner whose connection fails does
not answer late but simply does not answer. It is reliable without reducing to
correctness: response order is essentially uncorrelated with accuracy at $0.02$, only
faintly related to picking the eventual majority at $-0.05$, and weakly related to
self-report, with only two survey items significant at around $-0.11$ and $-0.13$
(Fig.~\ref{fig:splithalf}). It is not driven by poll size, correlating $0.14$ with a
learner's average poll size, and its reliability rises monotonically with polls per
learner, from 0.88 at five to 0.96 at thirty, so the ten-poll choice is conservative.

\begin{figure}[t]
\centering
\includegraphics[width=\columnwidth]{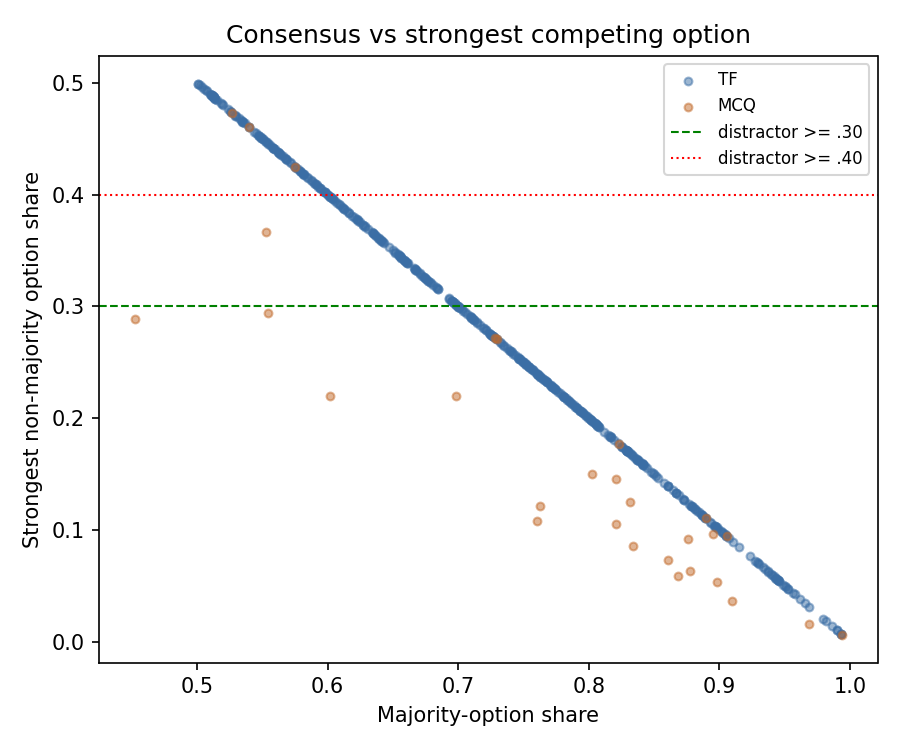}
\caption{Consensus versus strongest competing option. Each point is a poll,
positioned by its majority-option share and its second-option share. Items towards the lower right combine a weak majority with a large single competing block.}
\label{fig:scatter}
\end{figure}

\textit{A key-free map of cohort disagreement (RQ2).} Read without a key, the response distributions show a class that leans towards one
option but divides far from uniformly. Across the 505 cognitive fixed-choice items the
mean majority-option share is 0.727, median 0.732, with a long tail reaching 0.452.
The two types differ: the 476 true or false items average a majority share of 0.724
and normalised entropy 0.783, reflecting many close two-option splits, while the 29
multiple-choice items sit higher at 0.771 and much lower in entropy at 0.545, because
multiple-choice agreement, when it forms, tends to be near-complete. With only 29 such
items, we read these contrasts as descriptive rather than tested. Splits are common. Among the 476 true or false items, 105 (22.1 percent) are nearly even splits whose minority still draws at least 40 percent. Across all 505 items, 110 (21.8
percent) have a majority share below 0.60. Roughly one poll in five failed to reach even a clear majority,
taking 0.60 as the cutoff (Fig.~\ref{fig:scatter}).

\begin{figure}[t]
\centering
\includegraphics[width=\columnwidth]{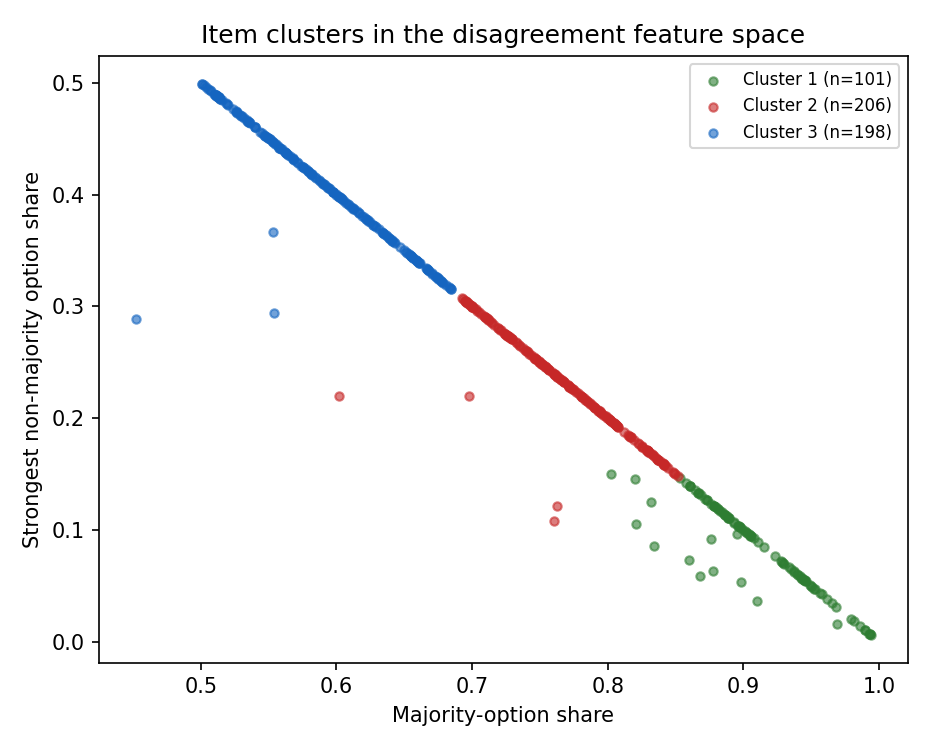}
\caption{The three item clusters in the disagreement feature space: a settled group
of high majority share and low entropy, a moderate group, and a strongly divided
group with a large competing block.}
\label{fig:clusters}
\end{figure}

Many divided items share a structure: a single competing option draws a large block
of the class. A second option draws at least 30 percent of the cohort on 211 of 505
items (41.8 percent) and at least 40 percent on 108, almost all true or false. Read
without a key, this flag ranks items by how strongly the class divides, not whether
the dividing block is right or wrong. Contested items are not spread evenly; they
concentrate in the more technical later material, with the largest counts in the
algorithms and linear-algebra sessions. These features sort the items into three
interpretive bands carried into the crossing (Fig.~\ref{fig:clusters}): a settled
group of 101 items (20.0 percent, centroid majority share 0.910, entropy 0.392), a
moderately divided group of 206 (40.8 percent, majority share 0.768), and a strongly
divided group of 198 (39.2 percent, majority share 0.591, second-option share 0.407,
entropy 0.964).

We reviewed the twenty most divided items by hand to separate genuine difficulty from item problems: twelve were genuine conceptual difficulty, six ambiguous wording, and two opinion prompts mislabelled as cognitive. The difficulty items are the encouraging case. One linear-algebra item asks whether the span of two vectors in three dimensions is always a line, cleanly dividing the learners who confuse a plane with a line; several others are fine-grained recall about specific session claims. The ambiguous items differ in kind, carrying absolute qualifiers or embedded negations that make the true-or-false direction hard to read. The two opinion items ask for a value judgment, so their near-even split reflects divided opinion, not a knowledge gap.

\textit{From contested to collectively wrong (RQ2, key-referenced).} Contestedness is key-free, so it cannot say whether a divided class is collectively
wrong or merely split on a hard question. A single key-referenced step separates the
two. On the 358 items with a settled, author-designated answer, the majority option is
wrong on 49 (13.7 percent), a collective misconception rather than mere division.
These errors sit where the key-free signal points: 41 of the 49 (84 percent) fall
among the contested items, and within the contested set the majority is wrong on 28
percent (41 of 146), rising to 32 percent (24 of 74) at the stricter 40 percent cut.
The rate climbs steadily with disagreement, from 4.2 percent on settled items above
0.80 to 35.0 percent on the most divided below 0.55. The key also recovers what
contestedness misses: eight of the 49 are not contested but confidently wrong, with up
to 93 percent of the class agreeing on the incorrect answer, so without a key they
read as settled. Key-free disagreement therefore flags most collective misconceptions
but not the confident ones, which only a key reveals.

\begin{figure}[t]
\centering
\includegraphics[width=\columnwidth]{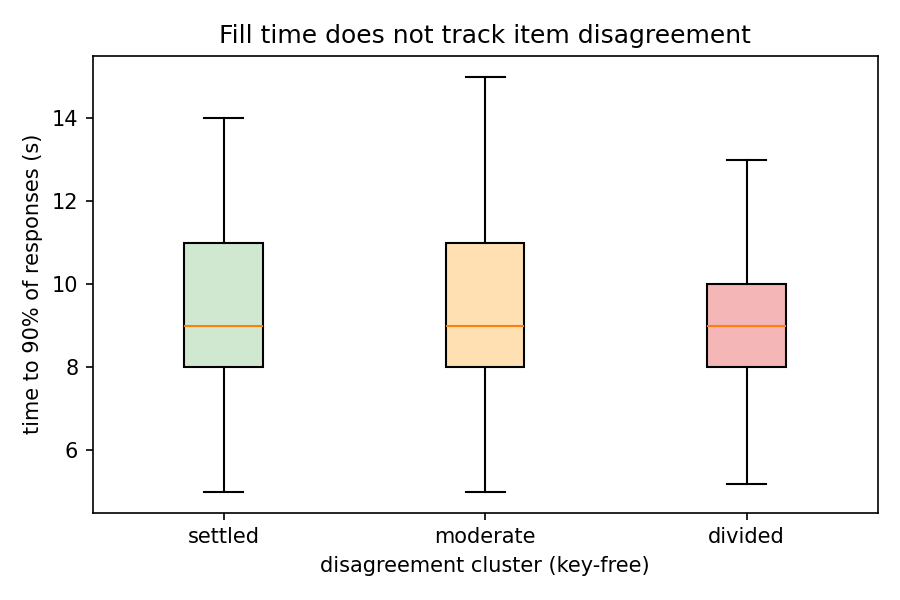}
\caption{Fill time by disagreement cluster. Items fill in the same short window
whether the cohort agrees or divides, so response timing and item disagreement are
independent key-free signals.}
\label{fig:fill}
\end{figure}

\textit{Crossing the two axes (RQ3).} The two readings are largely independent, and where they touch the interaction refines
rather than confirms the naive reading of the person axis. The item axis does not
predict the pace of the person axis: if divided items provoked hesitation they should
fill more slowly, but they do not. The median time to ninety percent of responses is 9
seconds in the settled, moderate, and divided clusters alike, a test of the three
fill-time distributions finds no difference (Kruskal-Wallis $p = 0.47$), and the trend
across ordered clusters is negligible (rank correlation $-0.04$). Read continuously,
time to ninety percent is essentially uncorrelated with a poll's entropy at $-0.07$,
holding after controlling for respondents. Polls fill in the same short window whether
the cohort agrees or divides, so fill speed is set by the shared pace of the
room, not item difficulty (Fig.~\ref{fig:fill}). Independence also holds at the learner
level: the correlation between response-order score and majority-choosing rate is
$-0.03$ on settled and $-0.01$ on divided items, neither significant, so how early a
learner answers says nothing about whether they end with the majority.

\begin{figure}[t]
\centering
\includegraphics[width=\columnwidth]{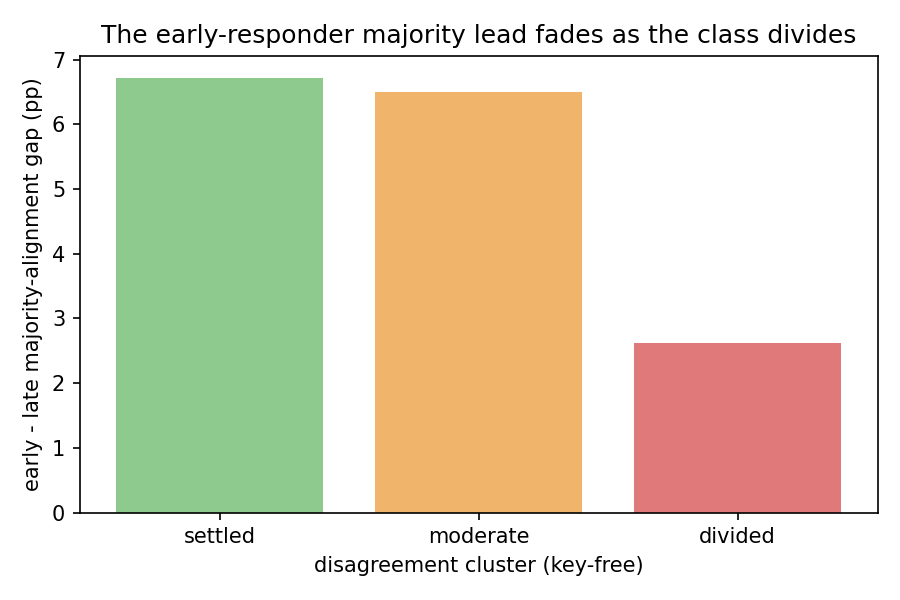}
\caption{The early-minus-late majority-alignment gap by disagreement cluster. The
early-responder lead is large on settled items and fades as the class divides.}
\label{fig:gap}
\end{figure}

The two axes meet at a single interpretable place. Averaged over all option-style
polls, learners in the earliest third match the eventual majority about 5 percentage
points more often than those in the latest third, a robust regularity on its own.
Crossing it with the item axis shows the effect concentrates where the class already
agrees (Fig.~\ref{fig:gap}): the early-minus-late alignment gap is 6.7 points on
settled items, positive on 86 percent of them, a comparable 6.5 on moderately divided
items, but only 2.6 on strongly divided items, positive on just 55 percent. Read
continuously, the gap declines as disagreement rises (rank correlation $-0.17$, $p =
1.5 \times 10^{-4}$). Early responders anticipate the majority most clearly where one
is already forming, and their lead largely dissolves where the cohort is genuinely
split. Contested items show the same pattern at finer grain: those who chose the large
competing block answer only marginally later than majority choosers, at a mean
response-order percentile of 0.508 against 0.495, detectable over more than a hundred
thousand votes but far too small to predict which side a learner takes.

\subsection{Discussion}
A live classroom poll is usually treated as a score of how many answered and how many were right. The same responses hold two other kinds of information that need no answer key. One is reading the learners who commit to an answer early and the ones who wait. The other is reading the questions where the class agrees and where it splits. This study asked whether these two readings can be trusted, and whether they carry the same information or different information.

The way a learner submits their answers, relative to classmates, is a stable habit. It stays with the learner across many polls and even within a single session, yet it has nothing to do with whether they answer correctly. It is therefore a real and consistent feature of how a student takes part, not an accident of the network or the clock, and not a hidden measure of ability. Reading the spread of answers without any key shows where a class is united and where it is divided. When a key is applied later, some of the divided questions turn out to be shared mistakes of the whole class, while others are simply hard, and the division itself points to most of those shared mistakes before any key is consulted.

Most importantly, the two readings are largely independent. How early a student answers says nothing about whether the class will divide, and a divided question does not slow the class down. The one place they meet is small and easily misread, since students who answer early tend to land on the eventual majority only when the class is already moving towards agreement. The early answering carries no foresight where the class is genuinely splitting. Being fast is not the same as being right, or leading the room.

For an instructor, a live poll can describe both the students and the questions in the moment, before any answer key exists, and the two descriptions are worth having together because each says something the other cannot. A teacher can see which questions genuinely divided the room and act on them while the lesson is still live, and can recognise a consistent pattern in how individual students take part. The same result carries a caution. Because early answering is a stable habit unrelated to correctness, it should not be rewarded as attentiveness or ability, and speed should never be read as a sign of being right. In this way, live polls can reveal valuable insights about the lessons being delivered and the common misconceptions students hold during a lecture.

\noindent\textbf{Limitations.} Two limitations qualify these readings. First, our person-axis measure is relative response order, not reaction time, so a stable per-learner network offset cannot be fully excluded, though its reach is bounded, since the order spreads learners widely and stays uncorrelated with correctness. Second, the setting is a single orientation programme dominated by two-option items, so the results are descriptive rather than causal and their generality awaits replication.

\section{Conclusion}
This paper has read a live classroom poll on two axes that need no key and asked how
they relate. Relative response order is a strongly stable learner signature: it
reproduces to a corrected split-half reliability of 0.92, holds at 0.87 within single
sessions, and spreads learners widely inside a synchronous window in which polls fill
in a median of thirteen seconds, yet it is uncorrelated with correctness (RQ1). Cohort
disagreement, read from the distribution alone, maps where the class divides and,
through a second-option flag, the 211 of 505 items where a large block settles on one
competing option, all without a key. The author-designated key, applied afterward,
shows the majority itself wrong on 49 of the 358 settled-answer items; the key-free
signal flags most of these collective misconceptions, though a divided class is more
often facing a hard question than a wrong answer (RQ2). Crossing the two axes is the
contribution: the readings are largely independent, since items do not fill more
slowly when more divided and response order does not predict which option wins, so
they describe learners and items separately, neither reducible to the other. They meet
at one place, where early responders anticipate the eventual majority only on
already-settled items, foreshadowing the outcome where a consensus is forming and
carrying no signal where the class is genuinely split (RQ3). These claims carry
limitations, which we set out in the Discussion.

Three directions follow. Joining the person axis to a recovered launch time would
convert relative response order into a reaction latency and connect it to the
psychometric response-time literature it now stands apart from. Widening the
multiple-choice arm and adding multi-rater coding of the difficulty-versus-ambiguity
distinction would strengthen the item map beyond the true-or-false-dominated set here.
And testing whether acting on either map, or their crossing, changes what a cohort
learns would move the account from description towards intervention.

\section*{Acknowledgement}
The per-item cognitive-level labels were prepared and checked by the authors. The answer key used only for
the key-referenced interpretation is the correct option designated by each question's author when the poll was created.

\section*{References}
{\footnotesize
\setlength{\parindent}{-1.2em}%
\setlength{\leftskip}{1.2em}%
\setlength{\parskip}{3pt}%
Asch, S. E. (1956). Studies of independence and conformity: A minority of one against a unanimous majority. \textit{Psychological Monographs, 70}(9), 1-70. \url{https://doi.org/10.1037/h0093718} \par
Black, P., \& Wiliam, D. (1998). Assessment and classroom learning. \textit{Assessment in Education: Principles, Policy and Practice, 5}(1), 7-74. \url{https://doi.org/10.1080/0969595980050102} \par
Caldwell, J. E. (2007). Clickers in the large classroom: Current research and best-practice tips. \textit{CBE-Life Sciences Education, 6}(1), 9-20. \url{https://doi.org/10.1187/cbe.06-12-0205} \par
Crocker, L., \& Algina, J. (1986). \textit{Introduction to classical and modern test theory}. Holt, Rinehart and Winston. \par
Crouch, C. H., \& Mazur, E. (2001). Peer instruction: Ten years of experience and results. \textit{American Journal of Physics, 69}(9), 970-977. \url{https://doi.org/10.1119/1.1374249} \par
De Boeck, P., \& Jeon, M. (2019). An overview of models for response times and processes in cognitive tests. \textit{Frontiers in Psychology, 10}, Article 102. \url{https://doi.org/10.3389/fpsyg.2019.00102} \par
Ebel, R. L., \& Frisbie, D. A. (1991). \textit{Essentials of educational measurement} (5th ed.). Prentice-Hall. \par
Gierl, M. J., Bulut, O., Guo, Q., \& Zhang, X. (2017). Developing, analyzing, and using distractors for multiple-choice tests in education: A comprehensive review. \textit{Review of Educational Research, 87}(6), 1082-1116. \url{https://doi.org/10.3102/0034654317726529} \par
Haladyna, T. M., \& Downing, S. M. (1989). A taxonomy of multiple-choice item-writing rules. \textit{Applied Measurement in Education, 2}(1), 37-50. \url{https://doi.org/10.1207/s15324818ame0201_3} \par
Haladyna, T. M., Downing, S. M., \& Rodriguez, M. C. (2002). A review of multiple-choice item-writing guidelines for classroom assessment. \textit{Applied Measurement in Education, 15}(3), 309-333. \url{https://doi.org/10.1207/s15324818ame1503_5} \par
Hestenes, D., Wells, M., \& Swackhamer, G. (1992). Force concept inventory. \textit{The Physics Teacher, 30}(3), 141-158. \url{https://doi.org/10.1119/1.2343497} \par
Hunsu, N. J., Adesope, O., \& Bayly, D. J. (2016). A meta-analysis of the effects of audience response systems (clicker-based technologies) on cognition and affect. \textit{Computers \& Education, 94}, 102-119. \url{https://doi.org/10.1016/j.compedu.2015.11.013} \par
Jensen, A. R. (2006). \textit{Clocking the mind: Mental chronometry and individual differences}. Elsevier. \par
Kay, R. H., \& LeSage, A. (2009). Examining the benefits and challenges of using audience response systems: A review of the literature. \textit{Computers \& Education, 53}(3), 819-827. \url{https://doi.org/10.1016/j.compedu.2009.05.001} \par
Lorenz, J., Rauhut, H., Schweitzer, F., \& Helbing, D. (2011). How social influence can undermine the wisdom of crowd effect. \textit{Proceedings of the National Academy of Sciences, 108}(22), 9020-9025. \url{https://doi.org/10.1073/pnas.1008636108} \par
Mazur, E. (1997). \textit{Peer instruction: A user's manual}. Prentice Hall. \par
Roediger, H. L., \& Karpicke, J. D. (2006). Test-enhanced learning: Taking memory tests improves long-term retention. \textit{Psychological Science, 17}(3), 249-255. \url{https://doi.org/10.1111/j.1467-9280.2006.01693.x} \par
Shrout, P. E., \& Fleiss, J. L. (1979). Intraclass correlations: Uses in assessing rater reliability. \textit{Psychological Bulletin, 86}(2), 420-428. \url{https://doi.org/10.1037/0033-2909.86.2.420} \par
Siemens, G., \& Baker, R. S. J. d. (2012). Learning analytics and educational data mining: Towards communication and collaboration. \textit{Proceedings of the 2nd International Conference on Learning Analytics and Knowledge}, 252-254. \url{https://doi.org/10.1145/2330601.2330661} \par
Smith, M. K., Wood, W. B., Adams, W. K., Wieman, C., Knight, J. K., Guild, N., \& Su, T. T. (2009). Why peer discussion improves student performance on in-class concept questions. \textit{Science, 323}(5910), 122-124. \url{https://doi.org/10.1126/science.1165919} \par
Surowiecki, J. (2004). \textit{The wisdom of crowds}. Doubleday. \par
van der Eijk, C. (2001). Measuring agreement in ordered rating scales. \textit{Quality \& Quantity, 35}(3), 325-341. \url{https://doi.org/10.1023/A:1010374114305} \par
van der Linden, W. J. (2007). A hierarchical framework for modeling speed and accuracy on test items. \textit{Psychometrika, 72}(3), 287-308. \url{https://doi.org/10.1007/s11336-006-1478-z} \par
Wang, A. I., \& Tahir, R. (2020). The effect of using Kahoot! for learning: A literature review. \textit{Computers \& Education, 149}, Article 103818. \url{https://doi.org/10.1016/j.compedu.2020.103818} \par
Wickelgren, W. A. (1977). Speed-accuracy tradeoff and information processing dynamics. \textit{Acta Psychologica, 41}(1), 67-85. \url{https://doi.org/10.1016/0001-6918(77)90012-9} \par
Wiliam, D. (2011). \textit{Embedded formative assessment}. Solution Tree Press. \par
Willis, G. B. (2005). \textit{Cognitive interviewing: A tool for improving questionnaire design}. Sage. \par
Wise, S. L., \& Kong, X. (2005). Response time effort: A new measure of examinee motivation in computer-based tests. \textit{Applied Measurement in Education, 18}(2), 163-183. \url{https://doi.org/10.1207/s15324818ame1802_2} \par
}

\end{document}